\documentclass[usenatbib]{mn2e}
\usepackage{graphicx}
\usepackage{amssymb}
\usepackage{longtable}
\usepackage{dpfloat}

\title[Metallicity Gradients at Large Radii]{Metallicity Gradients at Large Galactocentric Radii Using the Near-infrared Calcium Triplet}

\author[C. Foster et al.]{Caroline Foster$^{1}$, Robert N. Proctor$^{1}$, Duncan A. Forbes$^{1}$, Max Spolaor$^{1}$, \\\\
\normalfont{\LARGE{Philip F. Hopkins$^{2}$} and Jean P. Brodie$^{3}$}\\
$^{1}$Centre for Astrophysics \& Supercomputing, Swinburne University, Hawthorn, VIC 3122, Australia\\
$^{2}$ Department of Astronomy, University of California, Berkeley, CA 94720, USA\\
$^{3}$UCO/Lick Observatory, University of California, Santa Cruz, CA 95064, USA
}

\begin{document}

\maketitle

\begin{abstract}
We describe a new spectroscopic technique for measuring radial metallicity gradients out to large galactocentric radii. We use the DEIMOS multi-object spectrograph on the Keck telescope and the galaxy spectrum extraction technique of Proctor et al. (2009). We also make use of the metallicity sensitive near-infrared (NIR) Calcium\thinspace\textsc{ii} triplet (CaT) features together with single stellar population models to obtain metallicities. Our technique is applied as a pilot study to a sample of three relatively nearby ($\le30$ Mpc) intermediate-mass to massive early-type galaxies. Results are compared with previous literature inner region values and generally show good agreement. We also include a comparison with profiles from dissipational disk-disk major merger simulations. Based on our new extended metallicity gradients combined with other observational evidence and theoretical predictions, we discuss possible formation scenarios for the galaxies in our sample. The limitations of our new technique are also discussed.
\end{abstract}


\section{Introduction}

The study of the stellar population parameters of early-type galaxies (ETGs) is an invaluable tool to understand the star formation and enrichment history of such galaxies. Unfortunately, the determination of these parameters is greatly hindered by the age-metallicity degeneracy (Worthey 1994). This degeneracy affects photometric studies to a higher degree than spectroscopic studies. Because it is observationally expensive, typical spectroscopic studies of stellar populations concentrate on the central regions of galaxies reaching out to $\lesssim1 r_e$, where $r_e$ is the effective radius. This implies that typically at least 50 per cent of a galaxy's stellar mass is not surveyed spectroscopically, leaving an incomplete picture.

Moreover, feedback processes such as galactic winds from supernovae (SNe) or stellar mass loss, and active galactic nuclei (AGN) play an important role in the star formation history of galaxies. This is because feedback processes eject the gas that is required for the birth of new stars thereby shutting down star formation. Because metals are produced in stars, the shut down of star formation and thus feedback processes have a direct influence on the enrichment history of a galaxy. The efficiency of those feedback processes varies with galactocentric radius. On the one hand, galactic winds are more efficient at expelling gas at large galactocentric radii where the local potential well is shallower (Matteucci 1994; Martinelli, Matteucci \& Colafrancesco 1998). On the other other hand, due to their location and high efficiency, AGN can quench star formation at small and possibly out to large radii (e.g., Croton et al. 2006; Wang \& Kauffmann 2008). Therefore, radial profiles of stellar population parameters such as metallicity gradients can differ greatly from one galaxy to another depending on the relative importance of these feedback mechanisms and the formation mode (e.g., Spolaor et al. 2009). 

The processes involved in the formation and evolution of galaxies are still largely unknown despite tremendous efforts. The most popular scenarios have been the monolithic collapse and the hierarchical merging scenarios. In the classical monolithic collapse scenario (Larson 1974; Larson 1975; Carlberg 1984; Arimoto \& Yoshii 1987), galaxies form early via the \emph{dissipation} of large gas clouds, which induces the formation of the vast majority of the stars. Enriched gas from rapidly evolving stars in the outskirts sinks towards the centre thereby causing steep radial metallicity gradients (i.e., decreasing metallicity with galactocentric radius). The initial star formation period is then followed by quiescent evolution punctuated by few possible mergers and little subsequent star formation.

In contrast, in the classical hierarchical scenario (Kauffmann, White \& Guiderdoni 1993; Kauffmann \& Charlot 1998; De Lucia et al. 2006; De Lucia \& Blaizot 2007) galaxies form and evolve via the merging of sub-units. Star formation is then triggered by merger events throughout a galaxy's history. Under this paradigm, the assembly of galaxies is a \emph{continuous process} whereas the scale of the merger induced star formation episodes itself depends on the gas content of the progenitors involved. The high merger rates predicted by the hierarchical scenario should weaken or wash out metallicity gradients (White 1980, but see Di Matteo et al. 2009). However, as recently shown by the simulations of Hopkins et al. (2008, hereafter H08), merger induced central star formation due to \emph{dissipation} in gas-rich mergers can also create both shallow and steep radial metallicity gradients (see also Kobayashi 2004).

Over the years these two scenarios have been refined and modified to accommodate observations and their predictions accordingly updated. Recent literature has shown that modern versions of each of these scenarios can successfully reproduce elliptical-like objects (Kampakoglou, Trotta \& Silk 2008). At large radii the shape of the radial metallicity gradient predicted by these scenarios differs. The effect of dissipation, such as in a gas rich merger or under the monolithic collapse paradigm, yields steep gradients in the central regions of galaxies. However, at large radii the predictions for the slope of the metallicity gradient in gas-rich merger simulations tend to flatten (H08) while in the monolithic collapse simulations significant drops in metallicities are sometimes present (see Pipino, D'Ercole \& Matteucci 2008). In the absence of dissipation, such as in dry mergers, the predicted gradients are shallower and more or less constant with radius. Therefore, the change in slope of the metallicity gradient contains a `footprint' of the importance of dissipation in the star formation history of a galaxy.
 
The techniques used to derive metallicity gradients often suffer from the age-metallicity degeneracy. In their early study, Peletier \& Valentijn (1989) used radial colour gradients as probes for metallicity gradients assuming a constant age. Colour-magnitude diagrams of resolved individual stars have also been used to more robustly determine the metallicity of halo star populations for very nearby galaxies (e.g., Elson 1997; Harris \& Harris 2002; Rejkuba et al. 2005; Harris et al. 2007; Vlaji\'c, Bland-Hawthorn \& Freeman 2009). For unresolved stellar populations, even spectroscopic studies are not immune to the age-metallicity degeneracy as many authors have relied on only one or a few metal-lines measured from blue spectra (e.g., Spinrad et al. 1971; Cohen 1979; Gorgas, Efstathiou \& Arag\'on Salamanca 1990; Carollo, Danziger \& Buson 1993; Davies, Sadler \& Peletier 1993; Kobayashi \& Arimoto 1999; Ogando et al. 2005; Forbes, S\'anchez-Bl\'azquez \& Proctor 2005; Weijmans et al. 2009). In order to fully break the age-metallicity degeneracy, several studies (e.g., S\'anchez-Bl\'azquez et al. 2007; Brough et al. 2007; Spolaor et al. 2008b hereafter S08b) have used the technique of Proctor \& Sansom (2002), which uses $\chi^2$ fitting with as many spectral indices as possible, thereby yielding more accurate metallicity values. Unfortunately, even with this technique both the finite slit length and the faintness of a galaxy at large galactocentric radii prevented previous authors from efficiently measuring metallicity gradients to large radii especially in giant galaxies. Such studies have been recently compiled and expanded in Spolaor et al. (2009) to reveal an interesting inverse 'V-shape' relationship between the mass of galaxies and the slope of their inner metallicity gradient. At the low-mass end, the metallicity gradient goes from slightly positive slopes to more negative slopes as the mass increases. The relationship then reaches a turn-around at a mass corresponding to a velocity dispersion of roughly 140 km s$^{-1}$ after which galaxies with higher velocity dispersion (or mass) tend to have less negative metallicity gradients. The brightest cluster galaxies do not seem to fit this trend however and can scatter from null down to large negative gradient values.

In this pilot study, we extend the technique developed by Proctor et al. (2009, hereafter P09) to measure metallicity gradients out to large galactocentric radii for 3 giant ETGs (NGC 1407, NGC 2768 and NGC 4494). Following P09, we extract near-infrared (NIR) spectra of the galaxy halo light at large galactocentric radii using the DEIMOS multi-object spectrograph on Keck. The DEIMOS spectrograph is most efficient in the NIR where the Calcium\thinspace\textsc{ii} triplet (CaT) spectral feature dominates ($\sim8600$ \AA). The CaT has been shown to correlate with metallicity (e.g., Armandroff \& Zinn 1988; Diaz, Terlevich \& Terlevich 1989; Cenarro et al. 2001) with little age sensitivity (e.g., Diaz, Terlevich \& Terlevich 1989; Schiavon, Barbuy \& Bruzual 2000; Cole et al. 2004; Carrera et al. 2007; Vazdekis et al. 2007, hereafter V03). It is possible that different  Here we use the CaT together with the single stellar population (SSP) models of V03 to develop a new technique for deriving metallicity gradients using DEIMOS reaching as far out as $\sim 2r_e$. This technique can be applied to a large sample of galaxies and compared to the predictions of galaxy formation scenarios.

Our paper is divided as follows: in Section \ref{sec:sample} and \ref{sec:Data} we give a brief description of our sample galaxies and an overview of our data, respectively. In Section \ref{sec:Analysis} we explain the method used to extract metallicities and in Section \ref{sec:Results} we give our results. Finally, Sections \ref{sec:Discussion} and \ref{sec:Summary} contain a discussion of our results and a summary of this work, respectively.

\begin{table*}

\begin{tabular}{c c c c c c c c c c c}
\hline\hline
Galaxy & Hubble & P.A.& Axis ratio  & Distance & $r_{e}$ & $M_{B}$ & $M_{K}$ & Stellar mass & $V_{sys}$ & $\sigma_{0}$\\
 & Type & (deg) &(K band) & (Mpc) & (arcsec) & (mag) & (mag) & ($10^{11}M_{\odot}$) & (km s$^{-1}$) & (km s$^{-1}$)\\
 (1) & (2) & (3) & (4) & (5) & (6) & (7) & (8) & (9) & (10) & (11)\\
\hline
NGC 1407 & E0 & 60& 0.95 & 26.8 & 70 & -21.4 & -25.4 & 2.86 & 1782 & 272.5\\
NGC 2768 & S0$_{1/2}$ & 93& 0.46 & 20.8 & 64 & -20.9 & -24.6 & 1.37 & 1327 & 181.8\\
NGC 4494 & E1-2 & 173& 0.87 & 15.8 & 49 & -20.4 & -24.8 & 1.64 & 1335 & 150.2\\
\hline
\end{tabular}
\label{table:Sample}
\caption{Galaxy properties. Hubble types (column 2) are as per NASA/IPAC Extragalactic Database (NED).
Position angles and axis ratios (columns 3, 4) are from 2MASS.
Distances (column 5) are based on surface brightness fluctuations (Tonry et al. 2001) and include the distance moduli correction of Jensen et al (2003).
Effective radii (column 6) are taken from the Third Reference Catalogue of Bright Galaxies (RC3, de Vaucouleurs et al. 1991).
$B$- and $K$-band absolute magnitudes (columns 7, 8) are calculated from RC3 and 2MASS apparent magnitudes, respectively, and using the distances quoted in column 5.
Stellar masses (column 9) are calculated from the $K$-band magnitude of column 8 assuming a $M/L_K$ ratio corresponding to the V03 SSP of age 10 Gyrs and solar metallicity.
Systemic velocities (column 10) are from P09.
Central velocity dispersions (column 11) are as per Paturel et al. (2003).}
\end{table*}

\section{Sample}\label{sec:sample}
\begin{figure}
\begin{center}
\includegraphics[scale=0.25]{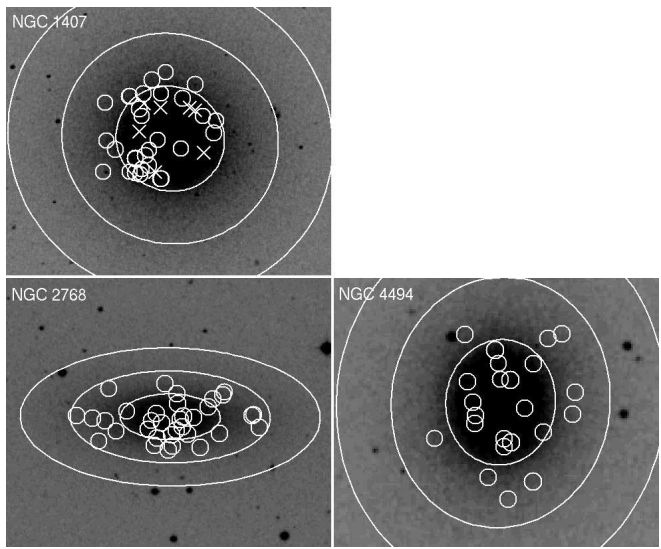}
\caption{Images of our sample galaxies from the DSS with overlaid position of the slits whose spectra have signal-to-noise ratio $\gtrsim8.5$ (small circles). Slits identified as crosses are discussed in Section \ref{sec:index}. Large circles correspond to 1, 2 and 3 effective radii. North and East are towards the top and left, respectively.}\label{fig:slits}
\end{center}
\end{figure}

A summary of the relevant properties of our sample galaxies is given in Table \ref{table:Sample}. Digitized Sky Survey (DSS) images are shown in Fig. \ref{fig:slits}. Below we give a brief overview of each galaxy. 

\subsection{NGC 1407}
NGC 1407 is the brightest group galaxy that dominates the Eridanus A group (Brough et al. 2006). It is a giant elliptical galaxy with a clear core-like luminosity profile (Spolaor et al. 2008a). It has been measured to have a uniform old age with a steep metallicity gradient and high $\alpha$-element ratio within one effective radius (S08b). In their study of the stellar kinematics, Spolaor et al. (2008a) found tentative signs of a kinematically decoupled core (KDC) at the centre of NGC 1407. However, as they point out this KDC detection may be spurious and due to a possible misalignment of the slit with the semi-major axis of the galaxy. Nevertheless, if confirmed, the presence of a KDC in the centre of NGC 1407 may be the signature of a merger history. Otherwise, it shows little sign of fine structure or recent disturbance.

\subsection{NGC 2768}
The literature is not unanimous with regards to the morphological classification of NGC 2768, which ranges from elliptical (de Vaucouleurs et al. 1991) to lenticular (Sandage, Tammann \& van den Bergh 1981; Sandage \& Bedke 1994). Evidence of rotation has been found by several authors (Fried \& Illingworth 1994; Emsellem et al. 2004; P09) as well as the presence of a centrally concentrated kinematic twist (McDermid et al. 2006). NGC 2768 is located at the centre of a loose group of which it is the brightest galaxy (Giuricin et al. 2000). No previous metallicity gradient measurement is available, however the measured central metallicity varies from $[Fe/H]\approx-0.2$ to $+0.3$ (see Howell 2005; Denic\'olo et al. 2005; Sil'Chenko 2006). The literature is also discrepant with respect to its central age with ages ranging between 8 and 15 Gyrs (e.g. Howell 2005; Denic\'olo et al. 2005; Sil'Chenko 2006) possibly because of different spatial sampling.

\subsection{NGC 4494}
NGC 4494 is classified as an elliptical galaxy located in the Coma I cloud (Forbes et al. 1996; Larsen et al. 2001). Its luminosity profile displays a central cusp (Lauer et al. 2007). It has a KDC (Bender et al. 1988), an intermediate central age of 6.7 Gyrs and a central metallicity of $[Fe/H]\approx+0.03$ (Denic\'olo et al. 2005). For these reasons, it is considered a good candidate for a possible gas-rich merger remnant (H08).

\section{Data}\label{sec:Data}

	\subsection{Acquisition}
Our main observing program was to obtain spectra for globular cluster systems around ETGs. A total of 5 ETGs have been observed so far (see P09 for details). From this initial sample, we selected 3 galaxies that had the greatest number of spectra and highest quality (signal-to-noise ratio).
	
Spectra were obtained using the DEIMOS spectrograph on the Keck telescope during the nights of 2006 November 19-21, 2007 November 12-14 and 2008 April 8. The seeing was good (typically $\sim 0.7$ arcsec). The 1200 l mm$^{-1}$ grating was used with a central wavelength of either 7500 \AA\space (2006 Nov.) or 7800 \AA\space (2007 Nov., 2008 Apr.). In every case, the setup allowed for the coverage of the CaT region ($\sim8400-8900$ \AA) with a resolution of $\Delta\lambda\sim1.5 \AA$ for the 1'' slit width. A total of 6, 2 and 3 multi-slit masks were observed for NGC 1407, NGC 2768 and NGC 4494, respectively. The typical total exposure time on each mask for NGC 1407 and NGC 2768 was 2 hours ($4\times30$ minutes exposures) and 1.5 hours ($3\times30$ minutes exposures) for NGC 4494. One mask for NGC 2768 was observed for 1 hour on two separate nights yielding two independent measurements. Fig. \ref{fig:slits} shows our selected galaxies together with the positions of the selected slits (see Section \ref{sec:reduction} for the selection process).

	\subsection{Reduction}\label{sec:reduction}

\begin{table}
\begin{center}
\begin{tabular}{ccc}
\hline\hline
Galaxy&Central passband&Continuum passbands\\
&(\AA)&(\AA)\\
\hline
NGC 1407&$8605.0-8695.5$&$8526.0-8536.0$\\
&&$8813.0-8822.0$\\
NGC 2768&$8605.0-8695.5$&$8478.0-8489.0$\\
&&$8813.0-8822.0$\\
NGC 4494&$8605.0-8695.5$&$8478.0-8489.0$\\
&&$8813.0-8822.0$\\
\hline
\end{tabular}
\caption{Sky scaling index definitions.}
\label{table:skyindex}
\end{center}
\end{table}	

The DEIMOS data were reduced using the \textsc{IDL} spec2d data reduction pipeline written for the DEEP2 galaxy survey. The pipeline performs both the flat fielding using internal quartz flats and the wavelength calibration using the ArKrNeXe arc lamps. Residual fringing is negligible due to instrumental design (e.g., Faber et al. 2008; Wirth et al. 2004). In addition to the sky subtracted globular cluster spectra, the pipeline produces several outputs, among which are the background (or `sky') spectra.

We used the technique of P09 to extract galaxy halo light spectra out of the background spectra. The background spectra are essentially the sum of both sky and galaxy light. P09 use an appropriately scaled `true' sky spectrum that is then subtracted from the background spectrum to extract the sky subtracted galaxy halo light spectrum. 

\begin{figure}
\begin{center}
\includegraphics[scale=.35,angle=-90]{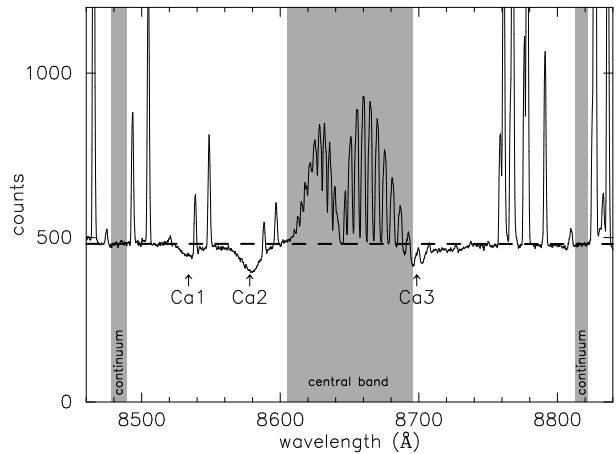}
\caption{Example of a background spectrum (i.e., galaxy light + sky) for NGC 2768. Labelled gray shaded regions highlight the continuum and central passbands of the sky scaling index definition. The dashed line shows the continuum estimate derived for this particular spectrum (the sky subtracted spectrum is shown in the top panel of Fig. \ref{fig:spectrum}). Positions of galaxy individual CaT features are labeled.}\label{fig:skyindex}
\end{center}
\end{figure}

In this work, we compute the scaling factor on the raw background spectra as the excess flux in the region 8605.0-8695.5\AA\space above a linear continuum determined from two carefully selected sidebands that avoid both strong skylines and galaxy spectral features for each galaxy (see Table \ref{table:skyindex}). Fig. \ref{fig:skyindex} shows an example background spectrum together with the definition of the sky scaling factor. As can be seen in Fig. \ref{fig:skyindex}, the recession velocity of NGC 2768 (and NGC 4494) causes the Ca3 line to be partly redshifted into the central passband of the sky index. This could introduce systematic errors when applying the sky subtraction. To test this, we use a slightly narrower central passband avoiding the Ca3 feature and find no noticeable changes to our results. Thus, because small variations in instrument resolution and wavelength solution across the mask could cause the minima between the individual skyline peaks to vary and introduce errors if the edge of the sideband is near a strong skyline we choose to use the same central band definition for all three galaxies in our sample. 

The `true' sky is derived from the normalized sum of several background spectra at large galactocentric radii (6-7$r_e$). Even at 6-7$r_e$, there is still some galaxy background light in our sky estimate. However, using a de Vaucouleurs' (1953) luminosity profile, we estimate that the galaxy light in these outer sky spectra is at most $6$ percent of the galaxy light in our science spectra. We use Monte Carlo methods to evaluate the accuracy of the sky subtraction at 9 signal-to-noise intervals ranging from 10 to 50 and find that the error in the final continuum level introduced by over/undersubtracting the sky is negligible compared to the noise (0.7 percent of the level of the noise for our lowest signal-to-noise ratio) if a linear continuum is assumed as is typical in this spectral region. The method described in P09 is similar to that of Norris et al. (2008) and Proctor et al. (2008), which employed background spectra from Gemini/GMOS and Keck/LRIS, respectively. For further information on the sky subtraction method used herein see P09. 

\begin{figure}
\begin{center}
\includegraphics[scale=.35,angle=-90]{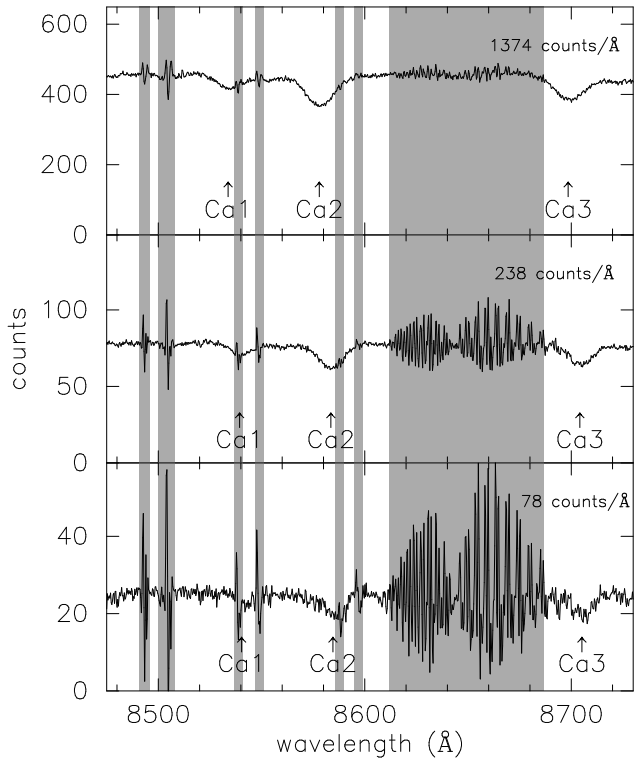}
\caption{Examples of a high (top panel), typical (middle panel) and low (bottom panel) signal-to-noise galaxy halo light spectra for NGC 2768. The raw spectra are shown as the black solid line. Gray highlighted regions show spectral ranges contaminated by skylines. Selected spectra have average number of counts per angstrom $\ge73$.}\label{fig:spectrum}
\end{center}
\end{figure}

Next, the galaxy halo light spectra were fitted using the \textsc{pPXF} code of Capellari \& Emsellem (2004) to extract halo kinematics out to large galactocentric radii ($\lesssim3r_{e}$). The kinematics are presented in P09. In Fig. \ref{fig:spectrum}, we show examples of the extracted galaxy halo light spectra and highlight regions that are still contaminated by skyline residuals. We find that approximately 10 per cent of the amplitude of these skyline residuals can be attributed to small variations of the wavelength solution across the mask. The remaining fraction is likely caused by the inherent complications associated with non-local sky subtraction due to variations of the sky spectrum over time and across the large field of view. As noted in P09, the strong skyline residuals are not significantly larger than the Poisson noise associated with them. Because of these skyline residuals, our method yields better results for galaxies with systemic recession velocities that do not cause the CaT features to be shifted into skyline dominated regions (i.e., 500 km s$^{-1}$ $\lesssim V_{sys} \lesssim 1400$ km s$^{-1}$ or 1700 km s$^{-1}$ $\lesssim V_{sys} \lesssim 2500$ km s$^{-1}$). Due to the inherent difficulties associated with flux calibrating multi-slit data our spectra are not flux calibrated.

Finally, we select our highest signal-to-noise spectra by removing any spectra with an average number of counts per angstrom $<73$. This roughly corresponds to a signal-to-noise cut of 8.5.

\section{Analysis}\label{sec:Analysis}

	\subsection{Index Measurements}\label{sec:index}
There are several CaT index definitions that apply to integrated light spectra. The choice of index definition is somewhat dependent on the purpose one desires to fulfill. Indeed, Armandroff \& Zinn (1988, hereafter AZ88) used narrow central passbands for the CaT to determine an empirical conversion between their CaT index and metallicities using the integrated light spectra of Galactic globular clusters. A year later, Diaz, Terlevich \& Terlevich (1989, hereafter DTT89) defined a much broader index, which they applied to the integrated light spectra of galaxies for which line indices are broadened due to their large velocity dispersion. The major drawback of this definition is that it uses the same continuum passbands for all three CaT lines and is therefore very sensitive to variations in the shape of the continuum. In the work of Cenarro et al. (2001, hereafter C01), this is circumvented by the possibility of using an arbitrary number of continuum passbands. Also, in order to reduce the effects of skylines and other non-Poisson noise during the CaT index measurement, the method of C01 weighs each pixel according to its variance. Moreover, the three features that constitute the CaT (i.e., Ca1, Ca2 and Ca3) have different relative depths with the bluest (Ca1) being the weakest. For this reason, some authors have decided to give varying weights to the different CaT features or to remove the Ca1 feature altogether from their CaT index definition (e.g. Armandroff \& da Costa 1991; Rutledge et al. 1997; Koch et al. 2006; Koch et al. 2008).

\begin{figure}
\begin{center}
\includegraphics[scale=.35,angle=-90]{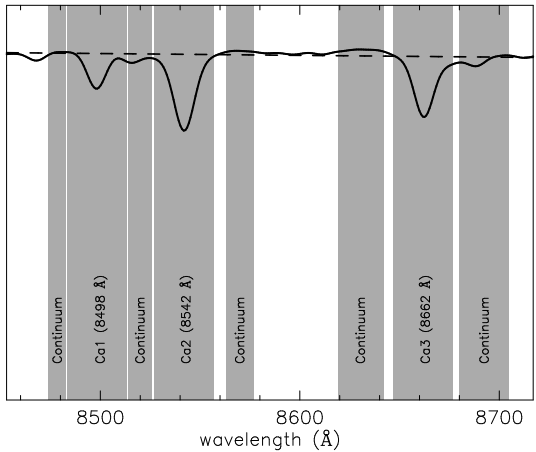}
\caption{Representation of the CaT definition used in this work. The solid line spectrum is a V03 SSP model with $[Fe/H]=-0.68$ and age of 13 Gyrs broadened to 150 km s$^{-1}$ dispersion to mimic the effect of a typical galaxy's velocity dispersion. Labelled gray shaded regions represent the CaT and continuum passbands. The dashed line shows the continuum level estimated using the method of C01.}\label{fig:definitionCaT}
\end{center}
\end{figure}

\begin{figure}
\begin{center}
\includegraphics[scale=.35,angle=-90]{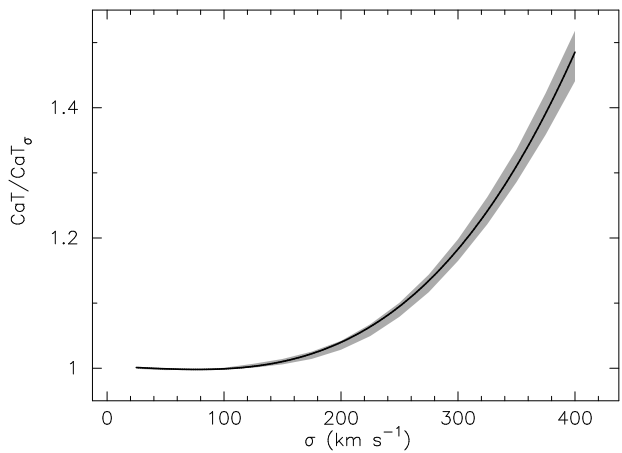}
\caption{Applied velocity dispersion index correction. The correction was obtained by comparing our index measured directly on the V03 SSP spectra of ages 13 Gyrs ($CaT$) and on the same spectra convolved with Gaussians of corresponding widths for a given velocity dispersion ($CaT_{\sigma}$). The shaded area shows the expected scatter in the correction caused by the different model metallicities. The solid line shows the best fit polynomial of third order through the median metallicity (i.e., $[Fe/H]\sim-0.28$) model points.}\label{fig:veldispcor}
\end{center}
\end{figure}

Because we are using velocity dispersion broadened galaxy spectra, we cannot use the narrow AZ88 central passbands definition or their empirical conversion. We thus choose to adopt the DTT89 index central passbands definition with the continuum determination and index measurement technique described in C01. The central passband definitions as well as the continuum passbands of our CaT index are given in Table \ref{table:IndexDefinition} and Fig. \ref{fig:definitionCaT}. We also employ a weighted sum such that:
\begin{equation}\label{eqn:CaT}
CaT=0.4\times Ca1+Ca2+Ca3.
\end{equation}
This is done in order to minimize the impact of the more uncertain Ca1 feature. The continuum passbands are chosen to uniformly cover the spectral range around the CaT features and in such a way as to avoid large spectral features and regions dominated by skyline residuals. The weights on the individual CaT features in Eq. \ref{eqn:CaT} are chosen to minimize the estimated errors on the CaT index values. Error estimates are computed using the background spectra (i.e., before sky subtraction) instead of fully propagated variance arrays, which are unavailable as our spectra are not processed within the data reduction pipeline. The background spectra provide a good estimate of the true variance since the bulk of the variance is caused by skylines, which in turn yields a robust error estimate. The technique described in Cardiel et al. (1998) and C01 (appendix A2) for generic indices is used. Finally, we apply a velocity dispersion correction in order to obtain CaT index values that are comparable to those measured at the models' dispersion. This correction is shown in Fig. \ref{fig:veldispcor}. For both NGC 2768 and NGC 4494, the velocity dispersions were taken from P09. However, the velocity dispersion profile of NGC 1407 measured by P09 reveals a sharp `spike' around $\sim0.9r_e$. The high velocity dispersion values measured from the corresponding slits (highlighted in white in Fig. \ref{fig:slits}) produce high CaT index values. These in turn yield highly deviant (unphysical) metallicities of up to $[Fe/H]\sim2.0$ around $r\sim0.9r_e$. The cause of this `spike' in the velocity dispersion profile is unknown. Because the required velocity dispersion correction is clearly too large, we adopt velocity dispersion values extrapolated through the `spike' by assuming a smoothly declining velocity dispersion profile as typical of elliptical galaxies.

The errors on the CaT index values are propagated accordingly and include errors in the respective velocity dispersion measurements. Because of the non-linear nature of the velocity dispersion correction, large velocity dispersions yield larger CaT index errors. Finally, we estimate that sky subtraction random errors contribute at most 0.001 \AA, which is minimal compared to our measured typical random errors. Larger systematics may arise from the skyline residuals as described in Section \ref{sec:reduction}. These are alleviated and partly accounted for via weighting according to the C01 index measurement method. In what follows we will refer to the CaT index measured using the above method as $CaT$.

In order to facilitate comparison with previous studies, we use the V03 SSPs with 13 Gyrs to derive a conversion between our $CaT$ and the similar CaT index defined by DTT89 ($CaT_{DTT}$). We obtain the following:
\begin{equation}
CaT_{DTT}=18.0-7.0\times CaT+0.86\times CaT^2,
\end{equation}
\noindent with a standard deviation of $\sigma=0.25$\AA.

\begin{table}
\centering
\begin{tabular}{c c c c}
\hline\hline
Feature & Weight & Central passband & Shared continuum\\
 & & (\AA) & passbands (\AA)\\
\hline
Ca1 & 0.4 & 8483.0-8513.0 & 8474.0-8483.0\\
Ca2 & 1.0 & 8527.0-8557.0 & 8514.0-8526.0\\
Ca3 & 1.0 & 8647.0-8677.0 & 8563.0-8577.0\\
 &  &  & 8619.0-8642.0\\
 &  &  & 8680.0-8705.0\\
\hline
\end{tabular}
\caption{CaT index definition.}
\label{table:IndexDefinition}
\end{table}

	\subsection{Conversion into metallicities}
We use the single stellar population (SSP) models of Vazdekis et al. (2003, hereafter V03) with the Kroupa (2001) initial mass function in order to convert our $CaT$ measurements into metallicities. There are two main reasons that motivated this choice. First of all, the V03 models provide spectral energy distributions (SEDs) at a resolution comparable to that of our data. Moreover, the V03 models show good agreement with Galactic globular cluster data (see fig. 14 of V03). We assume a constant age of 13 Gyrs to convert $CaT$ into metallicity. The predictions of the V03 models cover metallicities ranging from $-1.68\le[Fe/H]\le+0.20$ (corresponding to $4.1$ \AA\space $\lesssim CaT\lesssim6.2$ \AA). Outside this range, the behavior of the $CaT$ with metallicity is not constrained. This can be a serious issue particularly at high values of $CaT$ and $[Fe/H]$ where the conversion is steepest. For this reason and to avoid the introduction of uncertainties related to model extrapolation, we apply a hard boundary such that metallicities derived from data points that have measured $CaT\ge6.2$\AA\space are discarded and assigned the maximum metallicity (i.e., $[Fe/H]=0.2$ dex). These points are clearly identified in the plots that follow and are not used in subsequent analyses. Errors are propagated from the $CaT$ errors. Once again, the non-linearity of the derived conversion from $CaT$ into metallicity yields larger error bars at high metallicities. We also make use of rolling averages with radius as they are more robust against random fluctuations.

There could be other sources of uncertainty. First, there could be hot blue stars such as young, blue horizontal branch (BHB) or blue straggler (BS) stars contaminating the CaT with their predominant Paschen line features (see C01). Unfortunately, varying BHB morphologies and blue stragglers are not modeled by V03 and one has to worry about possible contamination by hot blue stars when converting $CaT$ into metallicities as 3 of the features in the Paschen series of hydrogen overlap with the 3 CaT lines. However, we visually inspected our spectra and find that there is no indication for the presence of a Paschen line at 8751\AA\space where it should be most easily seen.

\begin{figure}
\begin{center}
\includegraphics[scale=.35,angle=-90]{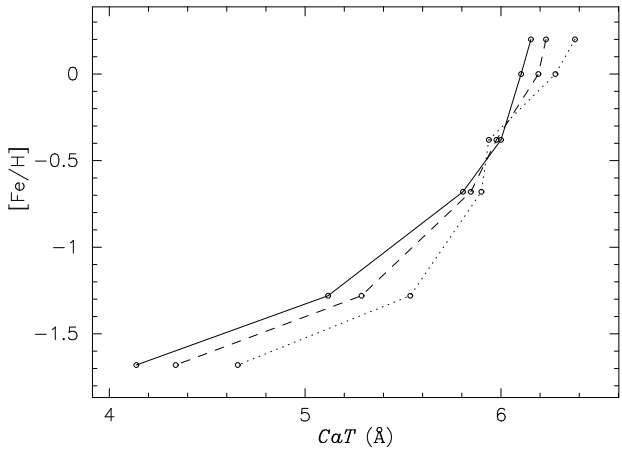}
\caption{Single stellar population model predictions for 13, 9 and 5 Gyrs (solid, dashed and dotted lines, respectively) from V03. The spectral resolution is $\Delta\lambda\sim1.5$\AA. The 5 Gyr model data are uncertain for metallicities $[Fe/H]=-1.28$ and --1.68.}\label{fig:models}
\end{center}
\end{figure}

Another source of uncertainty in our conversion into metallicities has to do with the effect of age on the CaT feature. Fortunately, the CaT is only minimally influenced by age effects for ages $\gtrsim2.5$ Gyrs (e.g., DTT89; Schiavon et al. 2000; V03; Cole et al. 2004; Carrera et al. 2007). This is in apparent contradiction with the predictions of the V03 models shown in Fig. \ref{fig:models}. On the other hand, it is likely that different CaT index definitions could yield different age dependencies as is often the case for other spectral features. Thus, it is conceivable that the apparent age trends seen in Fig. \ref{fig:models} may be influenced by our choice of index definition. Nevertheless, if we ignore the points corresponding to the 5 Gyrs models with metallicities $[Fe/H]=-1.68$ and --1.28 that V03 consider unreliable due to a lack of corresponding stars in the stellar library, the maximum error on the inferred metallicity induced by wrongly using a 13 Gyrs old SSP to estimate metallicity in a 5 Gyrs SSP is insignificant, as expected from observational studies. Therefore, because the inner parts of our galaxies are measured to be much older than 2.5 Gyrs, age effects should not strongly influence our inferred metallicities.

\section{Results}\label{sec:Results}

The spatial distribution of slits on the DEIMOS mask was optimized for the study of GCs. For this reason, the position from where our halo light spectra are extracted are distributed in a random manner around a given galaxy. In order to present our results in a way that is comparable to previous measurements (i.e., along the major or minor axes), we transform our galactocentric radii values into their spherical equivalent if the slit lies along the semi-major axis. To do this, we first define the effective change in right ascension and declination as 
\begin{eqnarray} 
\Delta \alpha&=&(\alpha_{slit}-\alpha_{galaxy})\cos\delta_{galaxy}, \mbox{ and} \\
\Delta \delta&=&\delta_{slit}-\delta_{galaxy}, \nonumber
\end{eqnarray}
\noindent respectively; where we have implicitly used the small-angle approximation. The subscripts `slit' and `galaxy' correspond to the position of the slit and the photometric centre of the galaxy in equatorial coordinates, respectively. Next we use the photometric axis ratio $(b/a)$, position angle $\phi$, and effective radius $r_e$ from the literature (see Table \ref{table:Sample}) to convert the 'true' distance to the centre of the galaxy into a pseudo major-axis distance. Or mathematically, we apply the following formula:

\footnotesize\begin{equation}
(r/r_e)=\frac{\sqrt{\left(\frac{\Delta\alpha'}{(b/a)}\right)^2+(\Delta\delta')^2}}{r_e}, 
\end{equation}
\normalsize

\noindent where $\Delta\alpha'=(\Delta\delta)\cos\phi + (\Delta\alpha)\sin\phi$, and $\Delta\delta'=-(\Delta\alpha)\cos\phi+(\Delta\delta)\sin\phi$. Assuming that there is no significant change in metallicity along isophotes, our results should be comparable to the literature's values for slits aligned with the semi-major axis. If this assumption is partly incorrect, it will introduce scatter in the metallicity at a given radius. In what follows, all galactocentric radii have been obtained with this method.

\begin{figure*}
\begin{center}
\includegraphics[scale=.55,angle=-90]{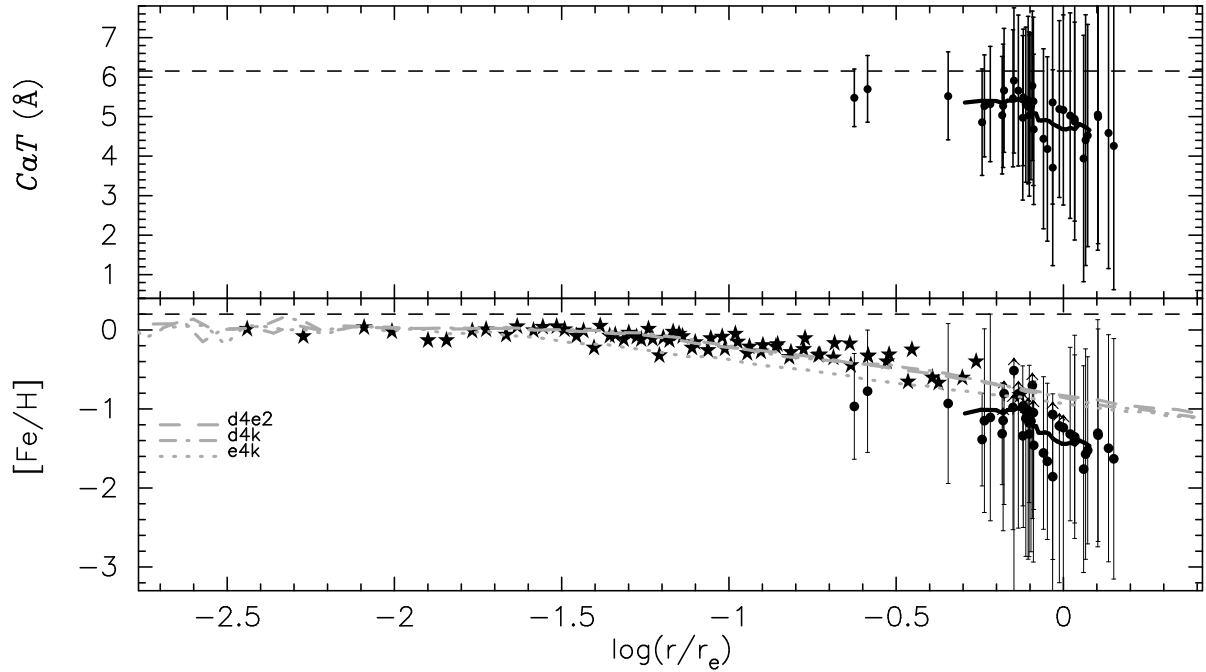}
\caption{$CaT$ (top panel) and metallicity (lower panel) gradients for NGC 1407. The stars are from the long slit data presented in S08b and the filled circles represent our data. Upper arrow error bars are used when the $CaT$ is beyond the upper limit (thin dashed lines) of the V03 SSP models. Thick solid lines are rolling averages. Profiles from the dissipational merger models of H08 are shown as thick gray lines (labels as per Table \ref{table:models}).}\label{fig:NGC1407}
\end{center}
\end{figure*}

We plot the rolling average using 8 to 10 points depending on the size of the dataset in Fig. \ref{fig:NGC1407}, \ref{fig:NGC2768} and  \ref{fig:NGC4494}. The rolling average gives a clearer indication of the actual shape of the metallicity gradient in our galaxies by eliminating the confusion caused by the scatter due to random errors on our data points. We find that the slope of the $CaT$ and metallicity gradients changes with galactocentric radius and thus cannot be appropriately described as linear in log-log space. This makes their objective quantification difficult.

The quantification of the metallicity gradient in the inner regions of galaxies using the slope of a fitted line in log-log space has been useful in previous works to quantify the steepness of the gradient and constrain theoretical models (e.g. Spolaor et al. 2009; Kobayashi \& Arimoto 1999). However, we do not fit linear relations to our metallicity gradients as this does not quantify them properly. Instead, we compare the entire profile with theoretical expectations (see Section \ref{sec:Discussion}).

\subsection{NGC 1407}

This galaxy is the only galaxy in our sample for which we have previous literature gradient values available to compare our results to. Unfortunately, it is not our best dataset in that the radial coverage is limited. Nevertheless, as seen in Fig. \ref{fig:NGC1407}, the general agreement of our $CaT$ measured metallicities with the metallicities measured from long-slit observations by S08b is reasonable. Although the individual values agree with the S08b data within the errors at similar galactocentric radii, the rolling average shows a slight offset in metallicity ($\sim0.3$ dex). The S08b data were obtained from optical spectra and calculated using the well tested $\chi^2$ fitting method of Lick indices of Proctor \& Sansom (2002). Considering that the present study may be sampling different stellar populations since 1) we sample a broader range of position angles, and 2) our wavelength range is much redder than S08b, this broad agreement is an \emph{indication} that the metallicities measured from the $CaT$ are reliable. Moreover, while we estimate that errors due to the lack of flux calibration of our spectra are negligible compared to the quoted errors, we cannot completely rule out the possibility that systematic offsets could be present.

Both the $CaT$ and derived metallicities reveal a steep gradient falling to $[Fe/H]\sim-1.8$ at $\log(r/r_e)\sim0.15$. Although the radial range covered here is limited, the $CaT$ and metallicity gradients show no clear sign of leveling off.

\subsection{NGC 2768}

\begin{figure}
\begin{center}
\includegraphics[scale=.35,angle=-90]{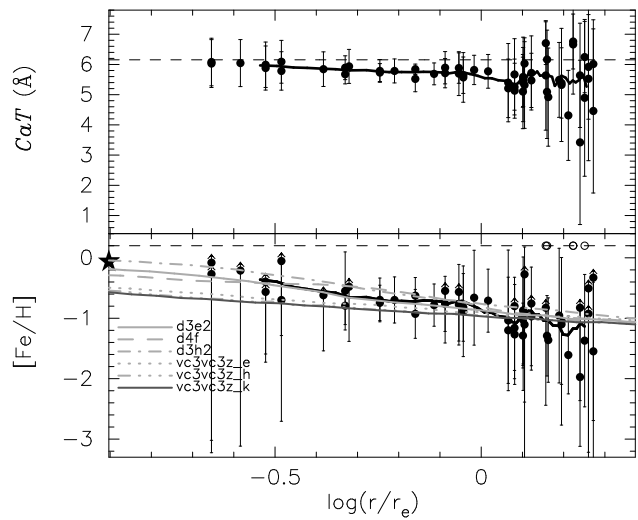}
\caption{$CaT$ (top panel) and metallicity (lower panel) gradients for NGC 2768. The star shows the central ($r\le r_e/8$) metallicity as measured by Howell (2005) and the circles represent our data. Hollow circles and upper arrow error bars are used when the $CaT$ is beyond the upper limit (thin dashed lines) of the V03 SSP models. Thick solid lines are rolling averages. Profiles from the dissipational merger models of H08 are shown as thick gray lines (labels as per Table \ref{table:models}).}\label{fig:NGC2768}
\end{center}
\end{figure}

As can be seen in Fig. \ref{fig:NGC2768}, the data for NGC 2768 cover a more extensive radial range than our NGC 1407 data. They also contain more individual measurements than for any other galaxy in our sample and exhibit smaller scatter. The central metallicity value measured by Howell (2005) is consistent with our innermost data points.
A visual inspection of Fig. \ref{fig:NGC2768} reveals a steep average metallicity gradient in the inner parts that steepens slightly before becoming shallower at $\log(r/r_e)\sim0.1$ around a metallicity of $[Fe/H]=-1.0$.

Although not inconsistent (i.e., still within the quoted errors), a few outer slits seem to exhibit $CaT$ values beyond the upper limit permitted by V03 (see Fig. \ref{fig:NGC2768}). Because of the uncertainties related to model extrapolation and especially given the steepness of the predicted relationship between $CaT$ and $[Fe/H]$ at high $CaT$ values, we consider these metallicity data unreliable.

\subsection{NGC 4494}
\begin{figure}
\begin{center}
\includegraphics[scale=.35,angle=-90]{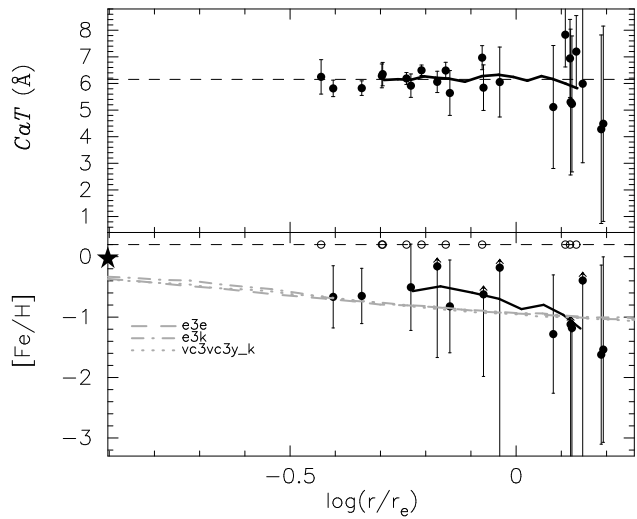}
\caption{$CaT$ (top panel) and metallicity (lower panel) gradients for NGC 4494. The star shows the central ($r\le r_e/8$) metallicity as measured by Denicol\'o et al. (2005) and the circles represent our data. Hollow circles and upper arrow error bars are used when the $CaT$ is beyond the upper limit (thin dashed lines) of the V03 SSP models. Thick solid lines are rolling averages. The metallicity gradient shown in the lower panel is fictitious (see text). Profiles from the dissipational merger models of H08 are shown as thick gray lines (labels as per Table \ref{table:models}).}\label{fig:NGC4494}
\end{center}
\end{figure}

Because shorter exposure times were used for NGC 4494, the number and spatial coverage of our measurements are much smaller than for the previous two galaxies. Despite the large scatter in our measured metallicity values, the agreement with the central estimate from Denic\'olo et al. (2005) is good. Interestingly, the data show no visible gradient in the raw $CaT$ measurements (see Fig. \ref{fig:NGC4494}) and the $CaT$ values seem to scatter about the upper limit of the V03 models. Moreover, the metallicity ``gradient'' seen in Fig. \ref{fig:NGC4494} is fictitious and caused by our exclusion of the randomly scattered points above the $CaT$ upper limit. Indeed, the majority of the measured metallicity points in NGC 4494 are very high (i.e., consistent with $[Fe/H]\gtrsim0.20$ dex) within the galactocentric radius probed and an accurate metallicity gradient could not be measured. For this reason, we cannot comment on the presence or absence of a metallicity gradient in this galaxy based on the current dataset.

It is clear that longer exposure times and more data are required to properly constrain the metallicity gradient in this galaxy at large galactocentric radii. Moreover, as for NGC 2768, the large $CaT$ values found indicate that the method presented herein may not be accurate at metallicities near solar and particularly beyond $[Fe/H]=+0.20$. Luckily, the negative radial metallicity gradients present in most galaxies mean that usually metallicities at large radii are sub-solar.

\section{Discussion}\label{sec:Discussion}

\begin{table}
\centering
\scriptsize
\begin{tabular}{c c c c c c}
\hline\hline
Simulation&Galaxy&Stellar mass&$\sigma_{0}$&$r_e$ & $f_{gas}$\\ 
 & & ($10^{11} M_{\odot}$) & (km s$^{-1}$) & (kpc) & \\
 (1) & (2) & (3) & (4) & (5) & (6) \\
\hline
d3e2&NGC 2768&1.00&158.9&4.02&0.18\\
d4e2&NGC 1407&2.51&231.2&5.26&0.20\\
d4f&NGC 2768&2.51&202.9&5.44&0.23\\
d3h2&NGC 2768&1.00&161.4&3.73&0.19\\
d4k&NGC 1407&2.51&226.1&4.93&0.23\\
e3e&NGC 4494&1.00&154.0&5.02&0.09\\
e3k&NGC 4494&1.00&154.9&5.03&0.11\\
e4k&NGC 1407&2.51&202.1&7.34&0.10 \\
vc3vc3y\_k&NGC 4494&1.00&154.0&4.46&0.14\\
vc3vc3z\_e&NGC 2768&1.00&145.1&5.90&0.03\\
vc3vc3z\_h&NGC 2768&1.00&139.4&6.14&0.03\\
vc3vc3z\_k&NGC 2768&1.00&140.8&5.86&0.04\\
\hline
\end{tabular}
\caption{Relevant properties of H08 dissipative major merger simulations (column 1) selected based on their matching of the luminosity profiles of our sample galaxies (column 2). The quoted stellar mass (column 3) is that of the remnant. Central velocity dispersions (column 4) are median line-of-sight values measured within 1 $r_e$ (as per column 5). $f_{gas}$ (column 6) is the mass fraction of the merging disks in the form of cold gas just before the final merger.}
\label{table:models}
\end{table}

In this pilot study, we develop a new technique to measure metallicity gradients at large galactocentric radii by expanding on the technique presented in P09. Using the CaT and the SSP models of V03, we obtain spectroscopic metallicity measurements out to 1.4$r_e$, 1.9$r_e$ and 1.5$r_e$ for NGC 1407, NGC 2768 and NGC 4494, respectively. We find that the metallicity gradients are not well described as a straight line in log-log space as their slope varies with galactocentric radius. The slope variations themselves contain information about the processes involved in the galaxy's formation. 

In order to better understand our observed metallicity gradients and the processes involved in their formation, we compare our results to the predictions from the models of H08. In H08, suites of hydrodynamic simulations of dissipational disk-disk galaxy merger remnants are compared to the observed properties (sizes and surface brightness profiles) of individual elliptical galaxies. The simulations determine the enrichment (i.e. metallicity) self-consistently from star formation and include a prescription for AGN and supernova feedback. For the individual galaxies in our sample, the models producing a remnant whose \emph{simulated} surface brightness profile yield the best $\chi^2$ matches to a given galaxy's \emph{observed} surface brightness profile are selected (see Table \ref{table:models}) with no prior on the metallicity or mass. \emph{Thus the predicted H08 metallicity gradients are not scaled to match the observed data.} In Fig. \ref{fig:NGC1407}, \ref{fig:NGC2768} and \ref{fig:NGC4494}, we show the line-of-sight averaged and \emph{B}-band luminosity-weighted stellar metallicity profiles predicted 3 Gyrs after the merger of these selected models for NGC 1407, NGC 2768 and NGC 4494, respectively. The simulated metallicity profiles are not strongly sensitive to the chosen line-of-sight or age of the merger.

Although the model metallicities are determined self-consistently from stars formed in the merger simulations, progenitor disks at the start of the simulation must have some `initial' metallicity. The chosen initial metallicities of the progenitors will not change the \emph{central} predicted metallicity of the remnant significantly as it is dominated by the merger induced dissipational star formation. However, the initial metallicity of the progenitors will come to dominate the metallicity of the remnant at sufficiently large radii. This sets an effective metallicity `floor' at large radii. 

First we compare our results with the simulated systems from H08 where the progenitor disks are initialized to lie on the observed redshift $z=0$ mass-metallicity relation (e.g., Tremonti et al. 2004) with a uniform metallicity at all radii. We find that this predicts too high a metallicity at large radii in the remnants to be compared with the current dataset. In the simulations shown in Fig. \ref{fig:NGC1407}, \ref{fig:NGC2768} and \ref{fig:NGC4494}, we lower the metallicity of the progenitors by a factor of $\sim3$ (i.e., $\sim-0.48$ dex). Again the metallicities at $\le1r_e$ are unchanged, but those at large radii are reduced. Under the dissipational major-merger formation paradigm, this lowered `floor' suggests that 1) the remnant's metallicity increasingly reflects that of its progenitors' stars as the radius probed increases; and 2) since the galaxies in question have relatively early formation times ($z \ge$ 1 to 2), the appropriate `initial' mass-metallicity relation is not that observed today (i.e. $z=0$) but that at these redshifts, which yields lower metallicities (e.g., Lara-L\'opez et al. 2009).

From the H08 simulations, we find that the central luminosity profile of NGC 1407 is well matched by mergers of intermediate gas-richness with gas fractions ($f_{gas}$) at the time of merger $\sim10$\%, typical of $L^*$ galaxies. The theoretical expectation is a relatively smooth metallicity profile, similar to that observed for $\log(r/r_e)\lesssim-0.1$. Once again, the different profiles shown correspond to the three simulations that provide the best match to the observed surface brightness profile with no prior on the metallicity or mass of NGC 1407. Thus, the amount of dissipation needed to match the luminosity profile appears to also provide a good match to the metallicity gradients for $\log(r/r_e)\lesssim-0.1$. Beyond that radius however there is an observed steepening of the metallicity gradient, which is not predicted by the models. This could be an indication that either 1) the initial metallicity of the progenitors is still too high (i.e., the `floor' should be set lower initially); or 2) that a dissipative major merger remnant is not a good description of the formation and evolution of NGC 1407. Under both the monolithic collapse scenario and dissipative merger scenarios, an early formation and assembly is required. Indeed, as concluded by S08b, NGC 1407's steep metallicity gradient, uniform old age and smooth photometric profile are consistent with a formation in which early dissipation played a major role. Moreover, it is not clear that a disk-disk major merger with plausible progenitors at moderate redshift could reproduce such a gradient unless the progenitors themselves possessed strong metallicity gradients (see Di Matteo et al. 2009). Alternatively, and still consistent with a hierarchical merging scenario, it is possible that the outskirts of NGC 1407 were built up primarily from smaller shredded systems (i.e., multiple minor-mergers, see Naab, Johansson \& Ostriker 2009), which typically have lower metallicities.

The matching of the observed and simulated photometric profiles of H08 is more ambiguous for NGC 2768 because it contains a lot of dust and exhibits several fine structures such as filaments and a ring of ionized gas (Martel et al. 2004; Lauer et al. 2005). This causes ambiguity with respect to selecting the best fitting models. Indeed, H08 find reasonable matches with \emph{both} relatively gas-poor mergers ($f_{gas}\sim3-5$\%) to relatively gas-rich ($f_{gas}\sim20-30$\%) ones. Unfortunately, the metallicity profiles are not sufficiently accurate to break this degeneracy. Nevertheless, the rolling average suggests that the simulations of H08 are a reasonable match to the measured metallicity profile and particularly for the relatively gas-rich models (i.e., $f_{gas}\sim20-30$\%).

Unfortunately, the metallicity gradient for NGC 4494 is not well constrained by our data. The shorter exposure time has yielded fewer data points to constrain the gradient. Moreover, at large $CaT$ (and velocity dispersions), the inferred metallicities are more uncertain and may yield unreliable results especially in modest signal-to-noise data. For this reason, we could not detect and measure a reliable metallicity gradient in NGC 4494. On the other hand, our data suggest that the metallicity of NGC 4494 is very high (i.e., $[Fe/H]\gtrsim0.20$ dex) out to $log(r/r_e)\approx0.2$.

\section{Summary and Future Work}\label{sec:Summary}
We describe a new technique for obtaining radial metallicity gradients of galaxies out to large galactocentric radii using the DEIMOS multi-object spectrograph on the Keck telescope. We use the NIR CaT spectral feature and convert our $CaT$ index into metallicity with the use of the V03 SSP models. This new technique is then applied to three intermediate mass to massive ETGs as a pilot study. Our results agree well with previous literature inner values. We find that at large galactocentric radii our measured metallicity gradients are not well described with straight lines (in log-log space) and show significant variations with galactocentric radius. A comparison to theoretical models is used in order to interpret our metallicity gradients. 

We conclude that NGC 1407 is likely to have formed the bulk of its stars via dissipational processes. This is consistent with the monolithic collapse scenario. However, we also hypothesize that the low metallicity stellar populations probed in the outskirts of NGC 1407 could have been hierarchically assembled from smaller low-metallicity systems. For NGC 2768, the measured metallicity profile is well reproduced by the dissipative major merger models of H08. We were unable to measure the metallicity gradient in NGC 4494 due to both the lower quality of this dataset and the fact that the measured $CaT$ values scatter around the upper limit permitted by the models. Nevertheless, we are able to conclude that the metallicity of NGC 4494 as measured via the $CaT$ is $[Fe/H]\gtrsim0.20$ dex within the galactocentric radii probed. The limitations of our new technique as a function of signal-to-noise, metallicity and velocity dispersion are also discussed.

The main science goal of this dataset was to obtain spectra of GCs. In the future we plan to include dedicated galaxy halo background slits in order to increase the radial and azimuthal coverage. This will enable 2-dimensional mapping of the metallicity distribution similar to the works of Kuntschner et al. (2006) and Rawle et al. (2008) reaching out to larger radii due to the larger field of view of the DEIMOS spectrograph. The assumption that the distribution of the metallicity follows the isophotes, which was imposed here, could then be tested and its effect on our conclusions assessed.

\section*{Acknowledgements}
We thank the anonymous referee for his/her careful reading and for offering very useful and constructive comments. We also thank G. Hau, L. Spitler, T. Mendel, J. Strader and A. Romanowsky for useful discussions and comments. CF thanks the Anglo-Australian Observatory for financial support in the form of a graduate top-up scholarship. RNP and DAF thank the ARC for financial support. Support for PFH was provided by the Miller Institute for Basic Research in Science, University of California Berkeley. This material is based upon work supported by the National Science Foundation under Grant AST-0507729. The data presented herein were obtained at the W.M. Keck Observatory, which is operated as a scientific partnership among the California Institute of Technology, the University of California and the National Aeronautics and Space Administration. The Observatory was made possible by the generous financial support of the W.M. Keck Foundation. The analysis pipeline used to reduce the DEIMOS data was developed at UC Berkeley with support from NSF grant AST-0071048. We acknowledge the usage of the HyperLeda database (http://leda.univ-lyon1.fr) and of NASA/IPAC Extragalactic Database (NED), which is operated by the Jet Propulsion Laboratory, California Institute of Technology, under contract with the National Aeronautics and Space Administration.

\begin{appendix}
\section{Data tables}
This section is a compilation of the results plotted in Fig. \ref{fig:NGC1407}, \ref{fig:NGC2768} and \ref{fig:NGC4494}. We quote the maximum value of $[Fe/H]=+0.2$ when the measured $CaT$ is larger than the range covered by the V03 models. However, we wish to caution the reader that while the quoted $CaT$ values are reliable within the quoted errors, the accuracy of the metallicity estimates is unknown for these specific data points only.

\noindent\begin{table}
\begin{tabular}{c c c c c}
\hline\hline
$\alpha$ & $\delta$ & $r/r_e$ & $CaT$ & $[Fe/H]$  \\ 
 (hh:mm:ss) & (dd:mm:ss) & & (\AA) & (dex) \\
 (1) & (2) & (3) & (4) & (5)\\
\hline
03:40:15.01&-18:35:13.25&0.772&5.30$\pm$1.96&$-1.12\pm$1.71\\
03:40:14.99&-18:35:30.48&0.929&3.71$\pm$2.48&$-1.86\pm$1.01\\
03:40:14.54&-18:35:33.22&0.895&4.18$\pm$2.33&$-1.66\pm$0.95\\
03:40:12.72&-18:35:39.05&0.789&5.24$\pm$1.90&$-1.17\pm$1.66\\
03:40:12.19&-18:33:22.63&1.267&5.05$\pm$3.43&$-1.31\pm$1.40\\
03:40:14.08&-18:35:08.81&0.571&4.86$\pm$1.34&$-1.39\pm$0.55\\
03:40:13.41&-18:33:31.77&1.165&4.41$\pm$3.17&$-1.57\pm$1.30\\
03:40:08.82&-18:35:06.04&0.660&5.27$\pm$1.56&$-1.15\pm$1.36\\
03:40:09.53&-18:33:38.80&1.149&3.94$\pm$3.12&$-1.76\pm$1.27\\
03:40:14.58&-18:34:09.34&0.790&5.03$\pm$2.04&$-1.32\pm$0.83\\
03:40:13.20&-18:35:30.60&0.707&5.46$\pm$1.73&$-0.98\pm$1.51\\
03:40:12.62&-18:33:49.87&0.872&4.44$\pm$2.28&$-1.56\pm$0.93\\
03:40:15.44&-18:33:53.09&1.080&4.94$\pm$3.06&$-1.35\pm$1.25\\
03:40:12.63&-18:34:08.37&0.605&5.32$\pm$1.45&$-1.11\pm$1.27\\
03:40:17.65&-18:34:00.92&1.364&4.59$\pm$3.43&$-1.50\pm$1.40\\
03:40:10.88&-18:35:00.40&0.260&5.70$\pm$0.67&$-0.77\pm$0.74\\
03:40:14.97&-18:35:12.31&0.759&5.49$\pm$1.73&$-0.96\pm$1.51\\
03:40:14.14&-18:34:02.78&0.807&5.78$\pm$1.89&$-0.70\pm$1.65\\
03:40:14.19&-18:33:50.00&0.972&5.19$\pm$2.24&$-1.21\pm$1.95\\
03:40:14.36&-18:34:18.89&0.664&5.66$\pm$1.56&$-0.81\pm$1.37\\
03:40:13.76&-18:35:21.71&0.656&5.03$\pm$1.48&$-1.31\pm$0.61\\
03:40:12.96&-18:34:49.05&0.237&5.48$\pm$0.60&$-0.97\pm$0.63\\
03:40:14.42&-18:35:26.27&0.800&5.26$\pm$1.86&$-1.15\pm$1.62\\
03:40:17.87&-18:35:29.42&1.414&4.26$\pm$3.64&$-1.63\pm$1.49\\
03:40:17.55&-18:34:49.50&1.182&4.52$\pm$2.81&$-1.52\pm$1.15\\
03:40:15.50&-18:35:29.44&0.999&5.17$\pm$2.41&$-1.24\pm$2.10\\
03:40:16.76&-18:35:01.29&1.047&5.03$\pm$2.60&$-1.32\pm$1.06\\
03:40:15.43&-18:33:52.96&1.081&4.87$\pm$2.52&$-1.38\pm$1.03\\
03:40:12.18&-18:33:22.44&1.269&4.99$\pm$3.20&$-1.33\pm$1.31\\
03:40:14.58&-18:34:38.79&0.580&5.27$\pm$1.28&$-1.15\pm$1.13\\
03:40:07.91&-18:34:40.78&0.815&4.68$\pm$1.90&$-1.46\pm$0.78\\
03:40:13.73&-18:35:01.58&0.452&5.52$\pm$1.10&$-0.93\pm$0.98\\
03:40:07.68&-18:34:25.46&0.928&5.36$\pm$2.57&$-1.07\pm$2.24\\
03:40:08.87&-18:34:19.33&0.756&4.97$\pm$2.08&$-1.34\pm$0.85\\
03:40:10.69&-18:33:56.20&0.813&5.39$\pm$2.13&$-1.05\pm$1.86\\
03:40:12.69&-18:35:38.79&0.784&5.42$\pm$2.12&$-1.02\pm$1.85\\
03:40:09.53&-18:34:11.48&0.733&5.66$\pm$1.91&$-0.81\pm$1.66\\
03:40:10.02&-18:34:08.12&0.711&5.91$\pm$1.84&$-0.51\pm$2.85\\
\hline
\end{tabular}
\caption{Individual values for NGC 1407. Columns 1 and 2 give the position of the individual slits in right ascension and declination (J2000), respectively. The major axis equivalent radius is given in column 3. Measured values of the $CaT$ index and $[Fe/H]$ are shown in columns 4 and 5.}
\label{table:NGC1407}
\end{table}

\clearpage

\begin{table}
\begin{tabular}{c c c c c}
\hline\hline
$\alpha$ & $\delta$ & $r/r_e$ & $CaT$ & $[Fe/H]$  \\ 
 (hh:mm:ss) & (dd:mm:ss) & & (\AA) & (dex) \\
 (1) & (2) & (3) & (4) & (5)\\
\hline
09:11:39.89&+60:01:37.87&1.274&6.03$\pm$2.01&$-0.27\pm$3.15\\
09:11:39.89&+60:01:37.87&1.274&5.32$\pm$2.08&$-1.10\pm$1.25\\
09:11:23.22&+60:02:24.78&1.671&6.76$\pm$2.27&$+0.20^*$\\
09:11:23.22&+60:02:24.78&1.671&6.66$\pm$2.46&$+0.20^*$\\
09:11:33.41&+60:02:16.73&0.468&5.89$\pm$0.73&$-0.55\pm$0.61\\
09:11:33.41&+60:02:16.73&0.468&5.68$\pm$0.75&$-0.79\pm$0.28\\
09:11:22.20&+60:02:08.74&1.820&5.92$\pm$2.74&$-0.51\pm$2.68\\
09:11:22.20&+60:02:08.74&1.820&5.52$\pm$2.97&$-0.92\pm$2.36\\
09:11:28.83&+60:01:58.20&1.204&5.25$\pm$1.78&$-1.16\pm$0.81\\
09:11:28.83&+60:01:58.20&1.204&5.67$\pm$1.84&$-0.80\pm$1.03\\
09:11:30.37&+60:02:41.56&1.163&5.21$\pm$1.72&$-1.20\pm$0.84\\
09:11:30.37&+60:02:41.56&1.163&5.40$\pm$1.79&$-1.04\pm$1.13\\
09:11:38.85&+60:02:07.47&0.300&5.99$\pm$0.50&$-0.39\pm$1.16\\
09:11:38.85&+60:02:07.47&0.300&5.88$\pm$0.50&$-0.56\pm$1.13\\
09:11:36.25&+60:01:59.28&0.568&5.79$\pm$0.76&$-0.70\pm$0.33\\
09:11:36.25&+60:01:59.28&0.568&5.73$\pm$0.77&$-0.75\pm$0.26\\
09:11:37.41&+60:01:32.92&1.434&6.70$\pm$2.35&$+0.20^*$\\
09:11:37.41&+60:01:32.92&1.434&5.64$\pm$2.59&$-0.83\pm$1.58\\
09:11:35.69&+60:02:12.86&0.222&6.03$\pm$0.44&$-0.27\pm$2.92\\
09:11:35.69&+60:02:12.86&0.222&6.08$\pm$0.44&$-0.08\pm$2.90\\
09:11:41.10&+60:01:58.14&0.692&5.84$\pm$0.93&$-0.63\pm$0.27\\
09:11:41.10&+60:01:58.14&0.692&5.52$\pm$0.96&$-0.93\pm$0.37\\
09:11:34.31&+60:01:54.53&0.817&5.71$\pm$1.10&$-0.76\pm$0.42\\
09:11:34.31&+60:01:54.53&0.817&5.89$\pm$1.12&$-0.55\pm$0.83\\
09:11:39.82&+60:02:19.25&0.328&6.09$\pm$0.56&$-0.05\pm$2.62\\
09:11:39.82&+60:02:19.25&0.328&5.78$\pm$0.57&$-0.70\pm$0.55\\
09:11:31.05&+60:02:31.14&0.883&5.88$\pm$1.19&$-0.56\pm$0.78\\
09:11:31.05&+60:02:31.14&0.883&5.70$\pm$1.22&$-0.78\pm$0.63\\
09:11:36.75&+60:01:38.21&1.264&5.59$\pm$2.02&$-0.87\pm$0.85\\
09:11:36.75&+60:01:38.21&1.264&5.51$\pm$2.12&$-0.94\pm$1.22\\
09:11:53.36&+60:02:10.14&1.870&6.01$\pm$3.12&$-0.33\pm$5.40\\
09:11:53.36&+60:02:10.14&1.870&4.46$\pm$3.36&$-1.55\pm$1.11\\
09:11:50.72&+60:02:07.76&1.566&5.39$\pm$2.43&$-1.04\pm$1.01\\
09:11:50.72&+60:02:07.76&1.566&5.32$\pm$2.58&$-1.10\pm$1.63\\
09:11:49.34&+60:01:38.98&1.782&6.24$\pm$3.03&$+0.20^*$\\
09:11:49.34&+60:01:38.98&1.782&4.90$\pm$3.29&$-1.37\pm$1.06\\
09:11:48.53&+60:02:05.28&1.323&5.47$\pm$1.96&$-0.97\pm$0.78\\
09:11:48.53&+60:02:05.28&1.323&5.72$\pm$2.02&$-0.76\pm$1.07\\
09:11:48.18&+60:02:46.95&1.736&5.64$\pm$3.22&$-0.83\pm$1.50\\
09:11:48.18&+60:02:46.95&1.736&3.42$\pm$3.24&$-1.97\pm$1.11\\
09:11:46.50&+60:01:52.47&1.263&5.41$\pm$1.80&$-1.02\pm$0.72\\
09:11:46.50&+60:01:52.47&1.263&5.10$\pm$1.86&$-1.29\pm$0.50\\
09:11:44.88&+60:02:19.45&0.905&5.64$\pm$1.35&$-0.82\pm$0.52\\
09:11:44.88&+60:02:19.45&0.905&5.57$\pm$1.38&$-0.89\pm$0.72\\
09:11:28.70&+60:02:48.06&1.445&6.16$\pm$2.41&$+0.20^*$\\
09:11:31.97&+60:01:38.66&1.444&5.09$\pm$2.39&$-1.29\pm$0.63\\
09:11:28.47&+60:02:51.46&1.545&5.49$\pm$2.48&$-0.96\pm$1.08\\
09:11:35.29&+60:02:03.77&0.478&5.94$\pm$0.68&$-0.48\pm$0.87\\
09:11:38.70&+60:02:57.40&1.454&4.92$\pm$2.66&$-1.36\pm$0.66\\
09:11:39.46&+60:01:47.13&0.961&5.82$\pm$1.35&$-0.66\pm$0.75\\
09:11:37.06&+60:01:52.67&0.769&5.69$\pm$1.03&$-0.78\pm$0.31\\
09:11:40.64&+60:02:09.32&0.414&5.85$\pm$0.65&$-0.62\pm$0.89\\
09:11:35.19&+60:02:16.80&0.261&6.05$\pm$0.47&$-0.21\pm$2.87\\
09:11:36.26&+60:02:33.27&0.616&5.79$\pm$0.79&$-0.70\pm$0.35\\
09:11:36.49&+60:02:45.93&1.039&5.77$\pm$1.46&$-0.71\pm$0.49\\
09:11:28.83&+60:01:58.16&1.205&5.13$\pm$1.81&$-1.27\pm$0.79\\
09:11:23.56&+60:02:23.41&1.627&4.31$\pm$2.33&$-1.61\pm$0.61\\
\hline
\end{tabular}
\label{table:NGC2768}
\caption{Individual values for NGC 2768. The column descriptions are the same as in Fig. \ref{table:NGC1407}. Because a mask was observed twice independently on separate nights, some positions are assigned two separate measured values.}
\vspace{-0.05 in}
\scriptsize{
$^* CaT$ value reached beyond the range covered by the V03 SSP models.}
\end{table}

\begin{table}
\begin{tabular}{c c c c c}
\hline\hline
$\alpha$ & $\delta$ & $r/r_e$ & $CaT$ & $[Fe/H]$  \\ 
 (hh:mm:ss) & (dd:mm:ss) & & (\AA) & (dex) \\
 (1) & (2) & (3) & (4) & (5)\\
\hline
12:31:23.50&+25:45:58.76&0.669&6.06$\pm$0.89&$-0.16\pm$1.48\\
12:31:27.86&+25:46:02.64&1.316&6.94$\pm$2.36&$+0.20^*$\\
12:31:25.55&+25:46:13.94&0.572&6.19$\pm$0.80&$+0.20^*$\\
12:31:25.51&+25:46:19.77&0.505&6.28$\pm$0.68&$+0.20^*$\\
12:31:25.64&+25:46:30.60&0.507&6.35$\pm$0.70&$+0.20^*$\\
12:31:19.76&+25:46:37.36&1.358&7.20$\pm$2.32&$+0.20^*$\\
12:31:25.94&+25:46:46.71&0.699&6.49$\pm$0.92&$+0.20^*$\\
12:31:24.05&+25:46:48.18&0.370&6.25$\pm$0.49&$+0.20^*$\\
12:31:24.15&+25:47:00.18&0.618&6.50$\pm$0.79&$+0.20^*$\\
12:31:24.32&+25:47:10.83&0.841&6.97$\pm$1.07&$+0.20^*$\\
12:31:26.06&+25:47:23.55&1.285&7.83$\pm$2.11&$+0.20^*$\\
12:31:23.67&+25:45:14.29&1.559&4.49$\pm$3.83&$-1.54\pm$1.50\\
12:31:22.24&+25:45:28.30&1.404&5.99$\pm$3.20&$-0.39\pm$4.59\\
12:31:24.82&+25:45:31.82&1.209&5.12$\pm$2.53&$-1.28\pm$0.94\\
12:31:23.86&+25:45:55.41&0.713&5.64$\pm$1.12&$-0.82\pm$0.73\\
12:31:23.88&+25:46:01.66&0.585&5.92$\pm$0.87&$-0.51\pm$0.68\\
12:31:21.61&+25:46:06.63&0.920&6.05$\pm$1.60&$-0.18\pm$4.85\\
12:31:19.91&+25:46:20.47&1.329&5.23$\pm$2.73&$-1.18\pm$2.22\\
12:31:22.65&+25:46:25.04&0.456&5.83$\pm$0.71&$-0.65\pm$0.42\\
12:31:23.46&+25:46:47.38&0.394&5.82$\pm$0.70&$-0.66\pm$0.48\\
12:31:22.11&+25:46:59.87&0.847&5.84$\pm$1.19&$-0.62\pm$1.32\\
12:31:21.26&+25:47:19.61&1.320&5.30$\pm$2.93&$-1.12\pm$2.39\\
12:31:20.45&+25:47:23.12&1.543&4.28$\pm$3.68&$-1.62\pm$1.45\\
\hline
\end{tabular}
\label{table:NGC4494}
\caption{Individual values for NGC 4494. The column descriptions are the same as in Fig. \ref{table:NGC1407}.}
\vspace{-0.05 in}
\scriptsize{
$^*CaT$ value reached beyond the range covered by the V03 SSP models.}
\end{table}
\end{appendix}
\end{document}